\documentclass[sigconf,natbib=true, nonacm]{acmart}

\AtBeginDocument{%
  \providecommand\BibTeX{{%
    \normalfont B\kern-0.5em{\scshape i\kern-0.25em b}\kern-0.8em\TeX}}}

\setcopyright{acmlicensed}
\copyrightyear{2024}
\acmYear{2024}
\acmDOI{XXXXXXX.XXXXXXX}

\acmConference[ArXiv'24]{}{}{}

\acmISBN{978-1-4503-XXXX-X/18/06}

\usepackage{caption}
\usepackage{multirow}
\usepackage{bm}
\usepackage{footmisc}

\begin{document}

\title{Cross-channel Recommendation for Multi-channel Retail}

\author{Yijin Choi}
\orcid{0009-0000-2827-7296}
\authornote{Equal contribution}
\affiliation{%
  \institution{UNIST}
  \city{Ulsan}
  \country{Republic of Korea}
  \postcode{44919}
}
\email{yjchoi0315@unist.ac.kr}

\author{Jongkyung Shin}
\orcid{0000-0003-4208-4313}
\authornotemark[1]
\affiliation{%
  \institution{UNIST}
  \city{Ulsan}
  \country{Republic of Korea}
  \postcode{44919}
}
\email{shinjk1156@unist.ac.kr}

\author{Chiehyeon Lim}
\orcid{0000-0001-6112-9674}
\authornote{Corresponding author}
\affiliation{%
  \institution{UNIST}
  \city{Ulsan}
  \country{Republic of Korea}
  \postcode{44919}
}
\email{chlim@unist.ac.kr}

\renewcommand{\shortauthors}{Choi et al.}

\begin{abstract}
An increasing number of retailers are expanding their channels to the offline and online domains, transforming them into multi-channel retailers. This transition emphasizes the need for cross-channel recommendations. Given that each retail channel represents a separate domain with a unique context, this can be regarded as a cross-domain recommendation (CDR). However, existing studies on CDR did not address the scenarios where both users and items partially overlap across multi-retail channels which we define as "cross-channel retail recommendation (CCRR)". This paper introduces our original work on CCRR using a real-world dataset from a multi-channel retail store. Specifically, we study significant challenges in integrating user preferences across both channels and propose a novel model for CCRR using a channel-wise attention mechanism. We empirically validate our model's superiority in addressing CCRR over existing models. Finally, we offer implications for future research on CCRR, delving into our experiment results.
\end{abstract}

\ccsdesc[500]{Information systems~Recommender systems}

\keywords{Cross-channel Recommendation, Cross-channel Integration, Multi-channel Retail}
\maketitle

\section{Introduction} \label{sec1}
\begin{figure}[t]
\centering
\includegraphics[width=0.95\columnwidth]{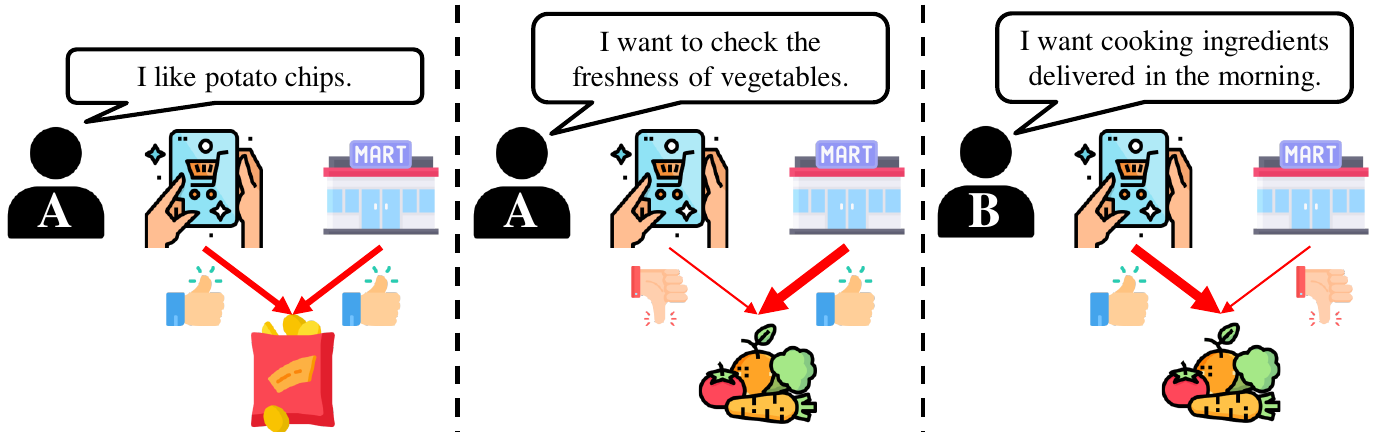}
\caption{Examples from multi-channel retail stores illustrate how purchase behaviors vary based on users, items, and channels.} 
\vspace{-10pt}
\label{fig1}
\end{figure}

To provide a seamless shopping experience, traditional retailers are transforming into multi-channel retailers \cite{cao2015impact, savastano2016going, timoumi2022cross}. For example, brick-and-mortar retailers (e.g., Walmart) and e-commerce dinosaurs (e.g., Amazon) are expanding their business to online and offline channels, respectively. This transition highlights the need for cross-channel recommender systems that provide personalized items within each channel based on users' purchase behaviors across channels to enhance competitiveness \cite{shi2020conceptualization, timoumi2022cross, balakrishnan2018product}. Meanwhile, existing studies in the retail domain have emphasized differences between online and offline environments \cite{luo2016online, mccabe2001information, shin2022recommendation} and have analyzed that users' purchasing behaviors in the two channels are different \cite{chu2010empirical, ariannezhad2021understanding, voorveld2016consumers}. As shown in Figure~\ref{fig1}, User A is willing to purchase potato chips from both channels but to purchase vegetables only from offline stores to check for freshness. User B prefers to buy vegetables from e-commerce for convenience. As such, the variation in purchase behaviors across channels by item, which differs by user, makes preference prediction difficult. This suggests that integrating user-item interactions from two distinct channels, known as cross-channel integration \cite{lee2010investigating, choi2020study}, is a significant challenge in providing the desired items to users within each channel.

\captionsetup[figure]{skip=5pt}

Considering the aforementioned, we introduce a new recommendation scenario, Cross-Channel Retail Recommendation (CCRR), which predicts user preferences for items tailored to each offline and online retail channel by utilizing interactions from both channels. Given that each channel involves a unique context for users' purchases, CCRR can be regarded as a cross-domain recommendation (CDR) that transfers knowledge from the source domain to the target domain to improve performance \cite{berkovsky2007cross, cantador2015cross}. However, CCRR is distinct from existing CDR studies that typically addressed scenarios where either only users overlap or only items overlap \cite{cremonesi2011cross, zang2022survey, chen2023toward}. In addition, they derive a \textit{similar} context through overlapping users or items across domains by considering domains with complementary relationships such as movie and book, or game and video \cite{guo2023disentangled, li2020ddtcdr, cao2023towards}. In contrast, CCRR is a scenario where users and items partially overlap across retail channels with interactions on each channel occurring in \textit{different} contexts and overlapping interactions present. Thus, when applied to CCRR scenarios, existing CDR methods can misinterpret user behavior on the source channel, misestimating item preferences on the target channel.

To address this issue, we propose $\mathrm{{C^{2}Rec}}$, a novel recommendation model equipped with a channel-wise attention mechanism to discern item preferences shared between channels and differentiated by channel. We perform joint learning on the tasks for recommendation and interaction classification. To the best of our knowledge, this study is the first attempt to define the CCRR scenario and to address the challenge that has not been covered in existing CDR studies. Our experiments on a real-world dataset demonstrate the superiority of $\mathrm{{C^{2}Rec}}$ over existing models. In addition, we offer implications for future research on CCRR.
\section{Dataset and Implication} \label{sec2}
We used sales histories from a real-world retail store that expanded from offline to online channels. This dataset was labeled to identify where each interaction occurred, and removed duplication. Detailed statistics for each channel are presented in Table~\ref{table1}.

We conducted several experiments to study user purchase behaviors by channel\footnote{The experiments utilize the same evaluation metrics described in Section \ref{sec4}.}. For each channel, BPR \cite{rendle2012bpr} was trained on all users' interactions in the training dataset and evaluated on test dataset from the same channel (self-match) and different channel (cross-match) for overlapping users. As shown in Table~\ref{table2}, cross-match, which simply transfers information between channels, performs worse than self-match in both channels. This suggests that users' item preferences in one channel are not maintained and differ in another channel, as found in \cite{ariannezhad2021understanding}. Additionally, we compared the results when item candidates were set to the entire item set (w/ purchased) and when they excluded items in the interactions of each user in the target channel's training data (w/o purchased). Note that due to duplication removal in each channel, items in the training data were excluded from the users' ground-truths for evaluation. The inferior performance for `w/ purchased' is caused because items highly preferred in source channel are already present in the user's interactions in target channel's training data, which hinders the recommendation of new items. This result and the existence of overlapping interactions across channels indicate that the user behaviors of purchasing the item regardless of channel and their preferences are shared between channels. Given these conflicting user purchase behaviors in multi-channel retail stores, it is important yet difficult to accurately estimate item preferences across different channels. Based on these findings, we classify users’ item preferences into two categories: (1) channel-specific item preference, which varies according to the channel, and (2) channel-shared item preference, which remains consistent across channels.

\captionsetup[table]{skip=5pt}
\begin{table}[t]
\renewcommand{\arraystretch}{1.3}
\centering
\caption{Statistics of the dataset.}
\label{table1}
\resizebox{\columnwidth}{!}{%
\begin{tabular}{ccccccc}
\hline
Channel & Users & Items & Interactions & Sparsity & Users overlap & Items overlap \\ \hline 
Offline & 18,326 & 5,194 & 1,062,119 & 98.88\% & \multirow{2}{*}{649} & \multirow{2}{*}{2,094} \\
Online & 3,687 & 2,190 & 68,452 & 99.15\% &  & \\ \hline
\end{tabular}%
}
\vspace{-5pt}
\end{table}

\captionsetup[table]{skip=5pt}
\begin{table}[t]
\renewcommand{\arraystretch}{1.3}
\centering
\footnotesize
\caption{Evaluation results for overlapping users.}
\label{table2}
\resizebox{\columnwidth}{!}{%
\begin{tabular}{ccccccc}
\hline
\multirow{2}{*}{Case type} & \multirow{2}{*}{Item candidates} & \multicolumn{2}{c}{Online} & & \multicolumn{2}{c}{Offline} \\ \cline{3-4} \cline{6-7} 
 & & HR@5 & NDCG@5 &  & HR@5 & NDCG@5 \\ \hline
Self-match & w/o purchased & \textbf{0.144} & \textbf{0.031} &  & \textbf{0.327} & \textbf{0.092} \\
Cross-match & w/o purchased & 0.083 & 0.013 &  & 0.107 & 0.019 \\
Cross-match & w/ purchased & 0.049 & 0.008 &  & 0.088 & 0.016 \\ \hline
\end{tabular}%
}\vspace{-10pt}
\end{table}

\section{Methodology} \label{sec3}
\subsection{Problem Definition}
Given the user-item interactions $\mathcal{I}=\{(u,v,c)|u\in\mathcal{U}, v\in\mathcal{V}, c \in\{off,on\}\}$ from the multi-channel retail store where $\mathcal{U}$ is the set of users, $\mathcal{V}$ is the set of items, and $c$ is the label that indicates where interaction occurs, the goal of CCRR is to recommend top-$k$ items for a user on each offline and online retail channel. Considering a multi-channel retailer's sales operations that share the same inventory, the entire $\mathcal{V}$ is assumed to be available in both channels. Since users and items overlap under CCRR, $\mathcal{U}^{off} \cap \mathcal{U}^{on} \neq \emptyset$ is satisfied, where $\mathcal{U}^{off}$ and $\mathcal{U}^{on}$ are the sets of users who purchased in offline and online channels, respectively. Furthermore, $\mathcal{V}^{off} \cap \mathcal{V}^{on} \neq \emptyset$ is satisfied where $\mathcal{V}^{off}$ and $\mathcal{V}^{on}$ are the set of items sold in offline and online channels, respectively. Under the CCRR scenario with overlapping interactions, the interactions are categorized into three types: those occurring exclusively in the offline channel ($\mathcal{I}^{off}$), exclusively in the online channel ($\mathcal{I}^{on}$), and those occurring across both channels ($\mathcal{I}^{both}$).

\captionsetup[figure]{skip=5pt}
\begin{figure}[t]
\centering
\includegraphics[width=0.99\columnwidth]{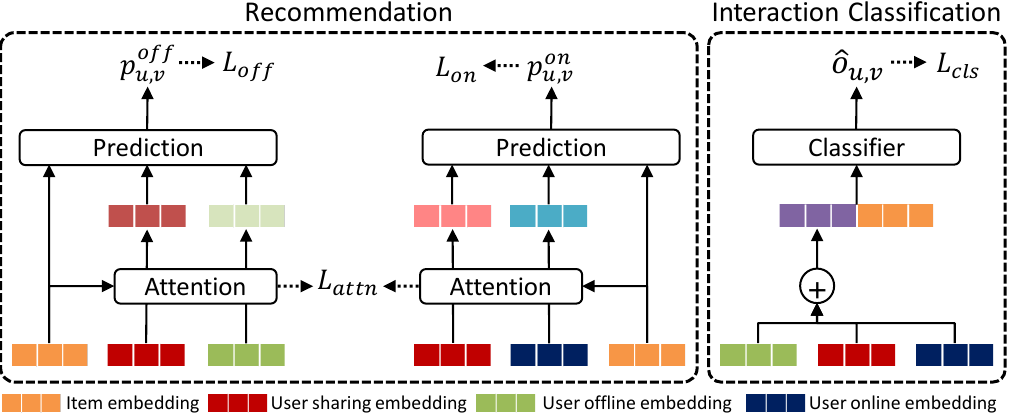} 
\caption{Overview of training mechanism of $\mathbf{C^{2}Rec}$.}
\label{fig2}
\vspace{-10pt}
\end{figure}

\subsection{Model Architecture}
To address the CCRR, we propose $\mathrm{{C^{2}Rec}}$ to capture both channel-shared and channel-specific item preferences across the channels. $\mathrm{{C^{2}Rec}}$ is composed of three user embedding layers, one item embedding layer, two attention blocks, two prediction blocks, and one interaction classifier. The training mechanism of $\mathrm{{C^{2}Rec}}$ is illustrated in Figure~\ref{fig2}. During training, $\mathrm{{C^{2}Rec}}$ simultaneously performs tasks for recommendation and interaction classification. Details of each component are described in the following subsections.

\subsubsection{User and Item Embedding}
We decompose user embedding into channel-sharing, offline, and online embeddings to reflect the distinct item preferences of users, based on findings from the experiments in Section \ref{sec2}. Accordingly, these embedding layers output a channel-sharing embedding $\bm{x_{u}^{sh}}$, an offline embedding $\bm{x_{u}^{off}}$, and an online embedding $\bm{x_{u}^{on}}$ for each user $u$, each having the same $d$ dimensionality. We describe these channel-specific embeddings, $\bm{x_{u}^{off}}$ and $\bm{x_{u}^{on}}$, as $\bm{x_{u}^{sp}}$. To this end, we introduce one item embedding layer to induce these decomposed user embeddings to have distinct representations for the same item. The item embedding layer outputs an item embedding $\bm{y_{v}}$ for each item $v$.

\subsubsection{Attention Mechanism}
Given that user interest varies depending on both item and channel, it is obvious that channel-shared and channel-specific preferences do not always have the same impact on the user's purchase behavior of an item in each channel. However, the channel-sharing and channel-specific embeddings for a user are consistent regardless of the item, failing to reflect the changing purchase behaviors across items. Therefore, we introduce a channel-wise attention mechanism to discern different preferences for each item in each channel. Given $\bm{x_{u}^{sh}}$, $\bm{x_{u}^{sp}}$, and $\bm{y_{v}}$, the attention mechanism aims to derive attention scores ($a_{u,v}^{sh}$, $a_{u,v}^{sp}$) indicating the relative importance of $\bm{x_{u}^{sh}}$ and $\bm{x_{u}^{sp}}$ for $\bm{y_{v}}$. The attention score of $\bm{x_{u}^{s}}$ for $s \in \{sh,sp\}$ is calculated as follows:
\begin{equation}
    a_{u,v}^{s}=\frac{exp \left( \frac{\bm{Q}(\bm{K^{s}})^{T}}{\sqrt{d'}} \right) }{\sum_{s' \in \{sh, sp\}} exp \left( \frac{\bm{Q}(\bm{K^{s'}})^{T}}{\sqrt{d'}} \right) },
\end{equation}
where $\bm{Q}=ReLU(h^{Q}(\bm{y_{v}}))$ and $\bm{K^{s}}=ReLU(h^{K}(\bm{x_{u}^{s}}))$ are $d'$ dimensional transformed representations of $\bm{y_{v}}$ and $\bm{x_{u}^{s}}$, respectively. $h^{Q}(\cdot)$ and $h^{K}(\cdot)$ are linear functions with a layer. Note that each attention block is configured independently, without sharing of layers across channels. Consequently, each representation of user $u$ depending on the item $v$ for channel $c$, $\bm{x_{u,v}^{c,s}}$, is obtained as $\bm{x_{u,v}^{c,s}} = a_{u,v}^{c,s}\cdot \bm{x_{u}^{s}}$ where $a_{u,v}^{c,s}$ is $a_{u,v}^{s}$ from attention block for channel $c$.

To ensure each attention score is informative for extracting users' distinct purchase behaviors across items and channels, we design an auxiliary objective to adjust the attention scores, considering the following. (1) The interactions in $\mathcal{I}^{off} \cup \mathcal{I}^{on}$ can be interpreted as having strong channel-specific characteristics in item purchases, indicating that users may or may not prefer certain items exclusively on specific channels. Conversely, the interactions in $\mathcal{I}^{both}$ demonstrate strong channel-sharing characteristics, suggesting that users have similar item preferences across channels. (2) The item purchase behaviors regardless of the channel and those that vary depending on the channel should have distinct attention score patterns. For item purchasing behavior with strong channel-specific characteristics, $a_{u,v}^{sp}$ should be higher than $a_{u,v}^{sh}$. To express strong channel-sharing characteristics,  $a_{u,v}^{sh}$ should be higher than $a_{u,v}^{sp}$. Based on these considerations, we propose attention loss to enforce the magnitude of attention scores depending on the channel where the interaction occurred as follows:
\begin{equation}
    \small
    \begin{split}
        \mathcal{L}_{attn} = -\frac{1}{|D|} \sum_{\scriptscriptstyle (u,v) \in D} \sum_{\scriptscriptstyle c \in \{off, on\}} (a_{u,v}^{c, sh} - (1-o_{u, v}^{s}))^{2} + (a_{u,v}^{c, sp} - o_{u, v}^{s})^{2},
    \end{split}
\end{equation}
where $D$ is the training dataset and $o_{u, v}^{s}$ denotes the indicator, which is a value of 1 if $(u, v) \in \mathcal{I}^{off} \cup \mathcal{I}^{on}$ and a value of 0 if $(u, v) \in \mathcal{I}^{both}$.

\subsubsection{Preference Prediction}
We make predictions for each $\bm{x_{u,v}^{c,s}}$ to comprehensively reflect user preferences by each characteristic separately. Similar with NeuMF \cite{he2017neural}, we calculate preference score in channel $c$ as follows:
\begin{equation}
p_{u,v}^{c} = \sigma(h^{sh}(\bm{x_{u,v}^{c,sh}} \odot \bm{y_{v}})+h^{c}(\bm{x_{u,v}^{c,sp}} \odot \bm{y_{v}})),
\end{equation}
where $\sigma$ denotes the sigmoid function and $\odot$ denotes the element-wise product. $h^{sh}(\cdot)$ and $h^{c}(\cdot)$ are the linear functions with one layer for channel-shared and channel-specific preferences, respectively. Then, the recommendation loss function for channel $c$ is defined to minimize the binary cross-entropy (BCE) as follows:
\begin{equation}
    \mathcal{L}_{c} = -\frac{1}{|D|} \sum_{(u,v) \in D} BCE(p_{u,v}^{c}, o_{u, v}^{c}),
\end{equation}
where $o_{u, v}^{c}$ is assigned as 1 indicating $(u, v)$ is occurred in channel $c$ and is assigned as 0 indicating it is not.

\subsubsection{Interaction Classification}
The attention mechanism extracts and trains the user's sharing and specific representations for each item. However, transforming decomposed embeddings depending on the item may inadvertently cause the trained representations to lose the user's unique characteristics. To preserve the unique characteristics of the user's purchase, we introduce an interaction classification task for incorporating the decomposed user embeddings. This task aims to classify the channel where an interaction occurred by reflecting user and item characteristics itself. For this, we integrated the decomposed user embeddings by the addition and concatenated the integrated user embedding $\bm{x_{u}}$ and item embedding $\bm{y_{v}}$ to represent interaction $\bm{i_{u, v}}=[\bm{x_{u}}; \bm{y_{v}}]$. Then, $\bm{i_{u, v}}$ feeds into a multi-label classifier with multiple linear layers. The classifier predicts $\bm{\hat{o}_{u, v}}$ where each element $\hat{o}_{u, v}^{c}$ indicates the probability of the interaction $(u, v)$ belonging to channel $c$. The loss for this classification task is calculated as follows:
\begin{equation}
    \mathcal{L}_{cls} = -\frac{1}{|D|} \sum_{(u,v) \in D} \sum_{c \in \{off, on\}} BCE(\hat{o}_{u, v}^{c}, o_{u, v}^{c}).
\end{equation}

\subsubsection{Multi-task Learning}
$\mathrm{{C^{2}Rec}}$ is trained to jointly optimize multiple objectives: (1) to predict channel-wise preferences and (2) regulate attention scores for the recommendation task, and (3) to preserve the unique characteristics of the user for the interaction classification task. Therefore, the overall training loss for $\mathrm{{C^{2}Rec}}$ is calculated as follows:
\begin{equation}
    \mathcal{L} = \mathcal{L}_{on} + \mathcal{L}_{off} + \lambda_{cls}\mathcal{L}_{cls} + \lambda_{attn}\mathcal{L}_{attn},
\end{equation}
where $\lambda_{cls}$ and $\lambda_{attn}$ are hyperparameters used to adjust the weight for $\mathcal{L}^{cls}$ and $\mathcal{L}^{attn}$, respectively.

\captionsetup[table]{skip=5pt}
\begin{table*}[t]
\renewcommand{\arraystretch}{1.3}
\centering
\caption{Performance comparison of the proposed model with baselines where best results are boldfaced.}
\label{table3}
\resizebox{\textwidth}{!}{%
\begin{tabular}{ccccccccccccccc}
\hline
\multirow{2}{*}{\begin{tabular}[c]{@{}c@{}}Test\\ dataset\end{tabular}} & \multirow{2}{*}{Metric} & \multicolumn{3}{c}{Single-Domain Recommendation} & \multirow{2}{*}{} & \multicolumn{3}{c}{Integration} & \multirow{2}{*}{} & \multicolumn{4}{c}{Cross-Domain Recommendation} & \multirow{2}{*}{$\mathrm{{C^{2}Rec}}$} \\ \cline{3-5} \cline{7-9} \cline{11-14}
 &  & BPR & NeuMF & NFM &  & BPR & NeuMF & NFM &  & CoNet & EATNN & DDTCDR & ETL &  \\ \hline
\multirow{4}{*}{Online} & HR@5 & 0.142(\textpm 0.009) & 0.153(\textpm 0.024) & 0.145(\textpm 0.024) &  & 0.150(\textpm 0.004) & 0.143(\textpm 0.005) & 0.134(\textpm 0.013) &  & 0.074(\textpm 0.011) & 0.121(\textpm 0.009) & 0.144(\textpm 0.009) & 0.110(\textpm 0.029) & \textbf{0.155(\textpm 0.002)} \\
 & HR@10 & 0.237(\textpm 0.003) & 0.229(\textpm 0.010) & 0.215(\textpm 0.023) &  & 0.216(\textpm 0.006) & 0.197(\textpm 0.005) & 0.192(\textpm 0.013) &  & 0.145(\textpm 0.004) & 0.224(\textpm 0.002) & \textbf{0.242(\textpm 0.008)} & 0.228(\textpm 0.012) & 0.229(\textpm 0.002) \\
 & NDCG@5 & 0.037(\textpm 0.002) & 0.043(\textpm 0.004) & 0.042(\textpm 0.008) &  & 0.044(\textpm 0.002) & 0.046(\textpm 0.003) & 0.041(\textpm 0.004) &  & 0.019(\textpm 0.002) & 0.030(\textpm 0.003) & 0.044(\textpm 0.008) & 0.027(\textpm 0.007) & \textbf{0.048(\textpm 0.001)} \\
 & NDCG@10 & 0.050(\textpm 0.002) & 0.050(\textpm 0.003) & 0.047(\textpm 0.008) &  & 0.049(\textpm 0.006) & 0.049(\textpm 0.002) & 0.045(\textpm 0.003) &  & 0.026(\textpm 0.001) & 0.042(\textpm 0.002) & 0.055(\textpm 0.007) & 0.041(\textpm 0.004) & \textbf{0.055(\textpm 0.001)} \\ \hline
\multirow{4}{*}{Offline} & HR@5 & 0.328(\textpm 0.019) & 0.321(\textpm 0.018) & 0.323(\textpm 0.007) &  & 0.313(\textpm 0.011) & 0.307(\textpm 0.016) & 0.298(\textpm 0.015) &  & 0.341(\textpm 0.020) & 0.347(\textpm 0.003) & 0.324(\textpm 0.012) & 0.347(\textpm 0.005) & \textbf{0.357(\textpm 0.002)} \\
 & HR@10 & 0.446(\textpm 0.010) & 0.442(\textpm 0.007) & 0.436(\textpm 0.001) &  & 0.441(\textpm 0.010) & 0.437(\textpm 0.005) & 0.420(\textpm 0.003) &  & 0.455(\textpm 0.007) & \textbf{0.456(\textpm 0.003)} & 0.441(\textpm 0.009) & 0.447(\textpm 0.005) & 0.455(\textpm 0.001) \\
 & NDCG@5 & 0.090(\textpm 0.004) & 0.087(\textpm 0.004) & 0.086(\textpm 0.002) &  & 0.087(\textpm 0.002) & 0.082(\textpm 0.004) & 0.079(\textpm 0.004) &  & 0.092(\textpm 0.006) & 0.092(\textpm 0.007) & 0.088(\textpm 0.002) & 0.095(\textpm 0.002) & \textbf{0.099(\textpm 0.001)} \\
 & NDCG@10 & 0.085(\textpm 0.002) & 0.081(\textpm 0.002) & 0.079(\textpm 0.001) &  & 0.082(\textpm 0.002) & 0.078(\textpm 0.001) & 0.075(\textpm 0.002) &  & 0.086(\textpm 0.003) & 0.085(\textpm 0.001) & 0.082(\textpm 0.002) & 0.085(\textpm 0.001) & \textbf{0.089(\textpm 0.001)} \\ \hline
\end{tabular}%
}\vspace{-2pt}
\end{table*}

\section{Experiment} \label{sec4}
We conducted experiments on a real-world multi-channel retail store dataset to verify the effectiveness of the proposed model.

\subsection{Experiment Settings}

\subsubsection{Datasets}
To construct train, validation, and test datasets, we utilized the aforementioned dataset. We randomly divided in a 6:2:2 ratio of interactions for each offline-only user in $\mathcal{I}^{off}$, interactions for each online-only user in $\mathcal{I}^{on}$, and interactions for each overlapping user in $\mathcal{I}$. Half of the interactions for overlapping users, which are extracted from $\mathcal{I}^{both}$ and assigned as validation and test datasets, are regarded to have occurred in one channel, and the remaining half in another channel. Then, we integrated all interactions assigned as train data. The validation and test datasets were constructed separately for each channel. For efficient training, we integrated negative samples, made by randomly selecting items that the user did not purchase from both channels, into the training dataset.

\subsubsection{Baselines and Evaluation Metrics} 
We have selected three single-domain recommendation (SDR) and four CDR baselines to compare with our model: BPR, NeuMF, NFM \cite{he2017neural2}, CoNet \cite{hu2018conet}, EATNN \cite{chen2019efficient}, DDTCDR \cite{li2020ddtcdr}, and ETL \cite{chen2023toward}. In addition to training the SDR baselines using the datasets of each channel, we also trained those by integrating the training dataset of both channels, referred to as Integration. The CDR baselines without DDTCDR were slightly modified to meet scenarios involving both overlapping users and items. For CoNet, the items inputted for each domain have been unified as the same item. For EATNN and ETL, if users purchased from only one channel, calculations on the item purchases for the channel where users did not interact were ignored in model training and their purchase vectors were represented as zero vector, respectively. We used Hit Ratio (HR@$k$) and Normalized Discounted Cumulative Gain (NDCG@$k$) \cite{jarvelin2002cumulated} as evaluation metrics commonly used in top-$k$ recommendations. We set $k$ as 5 and 10. For determining the best model in terms of validation performance, we utilized NDCG@10 as the criterion.

\subsubsection{Implementation Details} 
We trained the model with $k$ set to 10 for 200 epochs using the Adam optimizer \cite{kingma2014adam}. We set $d$ to 128 and the batch size to 1024. Other hyperparameters were determined through grid search with the following candidates: $d' \in \{64, 128, 256\}$, interaction classifier layer dimension $\in\{64, 128 \}$, learning rate $\in\{0.0001, 0.0005, 0.001\}$, $\lambda_{cls}\in\{0.1, 0.3, 0.5\}$, and $\lambda_{attn}\in\{0.01, 0.05, 0.1, 0.3, 0.5\}$. Early stopping with a patience of 20 was applied based on validation performance. All experiments were conducted with five different random seeds, and the mean and standard deviation of the results were reported.

\subsection{Performance Comparison}
Table~\ref{table3} shows the overall performance of our model and baselines. Our model generally outperforms all SDR and CDR baselines on both channels. Integration, which merely incorporates training datasets from channels, was not effective in improving the performances of SDR baselines on both channels. While CoNet, EATNN, and ETL, which use all channel data, perform better for the offline channel, they are inferior to SDR baselines for the online channel. In contrast, DDTCDR which truncates larger data to fit the size for both channels, excels over SDR baselines in the online channel but not in the offline channel. While Integration does not show benefit and the superiority of CDR baselines over SDR baselines varies by channel, our model consistently outperforms in both channels. This demonstrates our model's effectiveness in capturing different user preferences on items across channels.

\subsection{Ablation Study}
We investigate the contribution of each component through an ablation study on variants of $\mathrm{{C^{2}Rec}}$: (a) without interaction classification, (b) without attention assigning a constant weight of 0.5 to each channel-specific and channel-sharing embedding, (c) without attention loss, and (d) without separation of sharing and specific representations on channel, following general attention mechanism \cite{vaswani2017attention}. The performance of variants is reported in Table~\ref{table4}. Overall, all of these components can improve recommendation performance. Comparing (a) and (b), we observe that attention is more powerful than classification. It highlights the effectiveness of assigning optimal weights to shared and specific characteristics to derive preferences for each item and channel. The results for (c) and (d) show that regularizing the weight of each characteristic and considering user preference for each characteristic are valid for improving recommendation accuracy. Moreover, these outperform the CDR baselines, demonstrating that the two main components contributed significantly to the performance improvement.

\captionsetup[table]{skip=5pt}
\begin{table}[t]
\renewcommand{\arraystretch}{1.3}
\centering
\caption{Ablation study for the proposed model.}
\label{table4}
\resizebox{\columnwidth}{!}{%
\begin{tabular}{cccccc}
\hline
\multirow{2}{*}{Model} & \multicolumn{2}{c}{Online} & & \multicolumn{2}{c}{Offline} \\ \cline{2-3} \cline{5-6} 
 & HR@5 & NDCG@5 &  & HR@5 & NDCG@5 \\ \hline
$\mathrm{{C^{2}Rec}}$ & \textbf{0.155(\textpm  0.002)} & \textbf{0.048(\textpm  0.001)} &  & \textbf{0.357(\textpm  0.002)} & \textbf{0.099(\textpm  0.001)} \\ \hline
\multicolumn{1}{l}{(a) w/o classification} & 0.142(\textpm  0.008) & 0.043(\textpm  0.003) &  & 0.345(\textpm  0.004) & 0.094(\textpm  0.002) \\
\multicolumn{1}{l}{(b) w/o attention} & 0.132(\textpm  0.002) & 0.039(\textpm  0.001) &  & 0.329(\textpm  0.004) & 0.090(\textpm  0.001) \\
\multicolumn{1}{l}{(c) w/o attention loss} & 0.149(\textpm  0.002) & 0.046(\textpm  0.000) &  & 0.349(\textpm  0.001) & 0.096(\textpm  0.000) \\
\multicolumn{1}{l}{(d) w/o separation} & 0.149(\textpm  0.005) & 0.046(\textpm  0.002) &  & 0.352(\textpm  0.004) & 0.096(\textpm  0.001) \\ \hline
\end{tabular}%
}
\vspace{-8pt}
\end{table}

\subsection{Further Discussion}
In this experiment, we found that distinguishing different channel contexts and considering channel-sharing and channel-specific user behaviors by item are effective in the CCRR scenario. The ineffectiveness of Integration suggests that without distinguishing contexts, the model confuses user preferences for the target channel. While our model and CDR baselines consider this, only our model simultaneously improves the performance across both channels. CoNet, DDTCDR, and ETL predict user preferences by incorporating behaviors across channels without accounting for differences between channel contexts, resulting in biased user preferences on specific channels. While our model and EATNN distinctly recognize channel-shared and channel-specific purchase behaviors, our model can better understand users' item preferences in each channel by capturing the influence of these behaviors at the item level to consider different channel contexts. Meanwhile, for the online channel at HR@10, DDTCDR which reduces large datasets outperformed ours and other baselines trained on significantly larger offline channel data than online channel data. This implies that injecting unrefined source channel data with a significant discrepancy in quantity compared to the target channel may not be beneficial for predicting user preferences in the target channel. Our overall findings suggest that future CCRR research should focus on (1) methods to more effectively reflect the unique context of each channel and (2) developing data filtering techniques to extract relevant information from the source channel for the target channel.

\section{Conclusion} 
We originally studied the CCRR where users and items overlap across channels. Whereas existing studies on CDR have focused on similar contexts across users and items, in CCRR scenarios, addressing different channel contexts is key to understanding user purchase behavior in each channel. Our model achieves this and shows superiority through a channel-wise attention mechanism that extracts user preferences for each item across channels. We believe that our work will offer valuable insights into recommendation problems for multi-channel and omni-channel retailers.

\bibliographystyle{ACM-Reference-Format}
\bibliography{reference}

\end{document}